\documentclass{aa}  
\usepackage{graphicx,caption,subcaption}
\usepackage{txfonts}
\bibpunct{(}{)}{;}{a}{}{,}

\begin{document}

   \title{ The influence of general-relativity effects, dynamical tides and collisions on planet-planet scattering close to the star}

   \titlerunning{P--P scattering}

   \author{F. Marzari
          \inst{1}
          \and
          M. Nagasawa\inst{2}
          }

   \institute{University of Padova, Department of Physics and Astronomy\\
              \email{marzari@pd.infn.it}
         \and
             Kurume University, School of Medicine, Department of Physics\\
             \email{nagasawa\_makiko@med.kurume-u.ac.jp}
             }

   \date{}

 
  \abstract
   {Planet--Planet scattering is an efficient and robust dynamical mechanism for producing eccentric exoplanets. Coupled to tidal interactions with the central star, it can also explain close--in giant planets on circularized and potentially misaligned orbits. }
   {We explore scattering events occurring close to the star and test if they can reproduce the main features of the observed orbital distribution of giant exoplanets on tight orbits. }
   {In our modeling we exploit a numerical integration code  based on the  Hermite algorithm and including the effects of general relativity, dynamical tides and two--body collisions.}
   {We find that P--P scattering events occurring in systems with three giant planets initially moving on circular orbits close to their star produce a population of planets similar to the presently observed one, including eccentric and misaligned close--in planets.  The contribution of tides and general relativity is relevant in determining the final outcome of the chaotic phase. }
   {Even if two--body collisions dominate the chaotic evolution of three planets in crossing orbits close to their star, the final distribution shows a significant number of planets on eccentric orbits.  The highly misaligned close--in giant planets are instead produced by systems where  the initial semi--major axis of the inner planet  was  around 0.2 au or beyond.}

   \keywords{giant planet formation 
               }

   \maketitle
%

\section{Introduction}

Planet--planet scattering (hereinafter P--P scattering) is a physical phenomenon that has been initially invoked to explain the large orbital eccentricities observed among extrasolar planets. According to the widely accepted core--accretion model for planet formation \citep{alibert2005}, planets are expected to form on almost circular orbits. The subsequent onset of orbital instability may lead to a period of violent dynamical evolution dominated by close encounters between the planets.   This phase ends only when one or more planets are  ejected from the system on hyperbolic trajectories,  a collision(s) occurs between two planets or a planet(s) impacts the host star. The initial configuration of the planetary system is permanently and dramatically altered and the surviving planets are left on stable
highly eccentric non–interacting orbits possibly inclined respect to the initial
orbital plane \citep{weiden-marza96,rasio-ford96,MW02,namouni07,CHATTERJEE08,juritre08,ray08,naga08,ray09,marzari10,naga11,beauge2012}. 

The onset of the chaotic behavior can occur   at different stages of evolution of a planetary system. In the early phases, when the gaseous disk is still present, convergent migration \citep{masset2001,Leepeale2002}, may lead to crossing orbits and extended chaotic evolution  \citep{marza-barute2010,lega2013}. However, the subsequent interaction with the residual gas of the disk may partly damp the acquired high orbital eccentricities of the surviving planets.  Dynamical instability may also be triggered by the gas dispersal or it may suddenly start even after billions of years \citep{chambers1996, marweiden2002,veras2013}. In this last case the planets must lay within the orbital stability limit \citep{gladman1993,donnison2006,marzari2014} which depends on the masses and initial orbital parameters of the planets. P--P scattering can also take place when an external perturbation acts on the planetary system like a stellar flyby \citep{shara2016} unless it occurs in the early phases of evolution of a stellar cluster when its perturbing effects are erased by the gaseous disk damping \citep{marpi2013}. 

In its classical formulation, the P--P scattering model predicts that the inner planet gains the energy of the escaping one(s) moving into a closer orbit which, however, has a semi--major axis only slightly smaller than the initial one.  However,  in the following times, the energy loss due to tidal
interactions with the central star might circularize its trajectory near the periastron distance \citep{barnes2008,naga08} giving a significant contribution to populate the class of Warm/Hot Jupiters, gas giant exoplanets with very short orbital periods. In this context, the term ''eccentricity migration'' has been coined to indicate a giant planet, likely formed at several astronomical units by core--accretion, injected on an eccentric orbit and migrating very close to the star by tidal interaction with it.  The eccentricity pumping may occur either because of planet--planet scattering or even for different dynamical paths like Kozai interactions with an external massive companion \citep{fabry2007,naoz2011}. By comparing the distribution of the minimum separation of Hot Jupiters with their planet--star Roche separation \cite{nelson2017} have deduced that about 85\% of presently known Hot--Jupiters are possibly the outcome of eccentric migration while the others may be ascribed to disk--driven migration \citep{gold1980}. 

In a recent paper, \cite{petrovich2014} have argued that P--P scattering events occurring close to the star cannot excite high eccentricities or inclinations in giant planets since they mostly end up  in mutual collisions.  As a consequence, they cannot explain the observed eccentric and potentially misaligned Hot/Warm Jupiters orbiting, for example, within 0.1 au.  On this basis, they rule out the scenario where P--P scattering occurs after the planets are driven into close orbits by tidal interaction with the disk.  In alternative,  they suggest that planet instability takes place  far from the central star and the planets subsequently  migrate on eccentric and inclined orbits  till they get close to the star by interacting with the circumstellar disk. The occurrence of P--P scattering in presence of the  gaseous disk and far from the star has been studied with hydrodynamical simulations by \citep{marbaru2010,legamorb2013}. These simulations,  in addition to those performed by  \cite{marnel2009} for giant planets and by \cite{kleynelson2012} for smaller planets,  show that any initial high eccentricity and inclination of planets embedded in a disk  are rapidly damped on a timescale of the order of a few  $10^3$ yr, almost independently of the initial semi--major axis  of the planet. This complicates the scenario described by  \cite{petrovich2014} where they assume that migration must shrink the semi--major axis of excited planets by at least 1--2 orders of  magnitude without damping the eccentricity or inclination in order to produce Hot/Warm Jupiters on eccentric and misaligned orbits. Jupiter--mass planets might  have some of their orbital eccentricity preserved by their interaction with the disk through Lindblad and corotation resonances \citep{sari2003,angelo2006,duffell2015}. However, the maximum eccentricity achieved by these mechanisms is small, on average of the order of the disk aspect ration $H/r$ and it is limited to a maximum of 0.1--0.15 (for the more massive planets) while, for larger initial values of eccentricity,  damping is confirmed in all simulations. Even assuming that planets close to their star can  be influenced by the disk eccentricity, it is anyway difficult to explain the high values observed among exoplanets.  

Motivated by the results of \cite{petrovich2014}, derived with a pure N--body code  which does not include  the tidal interaction of the planets with the star and the effects of general relativity, we have adopted a more complete model, derived by that used by \cite{naga08}, to simulate P--P scattering events occurring close to the star.  This model, already including the effects of dynamical tides, has been improved by implementing the contribution from general relativity (isotropic, parametrized post-Newtonian) which may affect not only the evolution of a single close planet but also the secular evolution of a pair of planets \citep{must2014}, and mutual collisions with merging. We then focused on three-planet
systems  in which we consider different initial distances from the star of the inner one ranging from 0.03 to 1 au.  According to population synthesis models based on  core--accretion  and illustrated in \citep{emme19}, the fraction  of multi--planet systems which form  with either 2 or 3 giant planets is approximately the same but the P--P scattering dynamics for three planets is more  diverse.  We have run a large number of simulations and compared the outcomes with observations to test the reliability of our model.  We have also explored the influence of general relativity and dynamical tides on the final configurations of the systems. 

In Section 2 we describe the numerical  model used to simulate the evolution of a large number of putative three--planet systems moving initially close to the star. In Section 3 the possible outcomes of the dynamical instability are described. In Section 4 we show that general relativity can significantly alter the secular evolution of a pair of planets close to the star. In Section 5 we show how a random initial population of three planets evolved through the chaotic phase can well match the observed distribution. Finally, in Section 6 we summarize and discuss our results.  

\section{The numerical model}
\label{numerical}

We model P--P scattering events for planetary systems initially made of three equal mass giant planets with $m = 1 M_J$. Population synthesis models have shown that these systems may be a common outcome of the initial phases of evolution of a circumstellar disk \citep{emme19}, at least as frequent as systems consisting of two giant planets. We randomly select the initial orbital parameters of the inner planet while the semi--major axis of the second and third planet are computed close to the stability limit so that instability occurs on a timescale typically shorter than $10^5$ yrs. This choice does not significantly affect the final orbital distribution of the planets after the chaotic phase as shown in previous models of P--P scattering 
\citep{weiden-marza96,
rasio-ford96,
MW02,namouni07,CHATTERJEE08,juritre08,ray08,naga08,ray09,marzari10,naga11,beauge2012} and it allows to reduce the computational burden. The orbital angles of all planets are generated randomly while the initial eccentricity and inclinations are kept small, as expected from planet formation models.   

In the numerical algorithm  we include the effects of tides on the planet, mutual collisions and general relativity. 
Modeling the tidal interaction between a planet and the star is a complex problem and many aspects of this interaction are not fully understood. In general, one should consider both equilibrium and dynamical tides. For static tides, dissipative and frictional processes within the planet remove energy from its orbit while for dynamical tides this occurs through the excitation and damping of oscillations (see ad example \cite{ivanov2004, ivanov2007,ogilvie2013,mathis2015}). 

In our model we include tides related to the fundamental modes and inertial modes of a planet in a state of pseudo-synchronization as described in \cite{ivanov2007}. The energy transfer associated with the fundamental modes is minimal and no angular momentum transfer occurs in the state of pseudo-synchronization.
The energy transfer associated with the fundamental modes (hereinafter $\Delta E_{\rm f}$) and inertial modes (hereinafter $\Delta E_{\rm in}$) excited in the planet are given by equation (A7) and equation (100) of \cite{ivanov2007}, respectively,  where we assume the typical coefficient 
$3 \times 10^{-3}$ even for a $1 R_J$ planet. 
In our simulations, the tidal effect  is modeled as impulsive changes in energy  given by $\Delta E_{\rm f}+\Delta E_{\rm in}$  to the planetary orbits at every periastron passage.

\begin{figure}[hpt]
\includegraphics[width=1.0 \hsize]{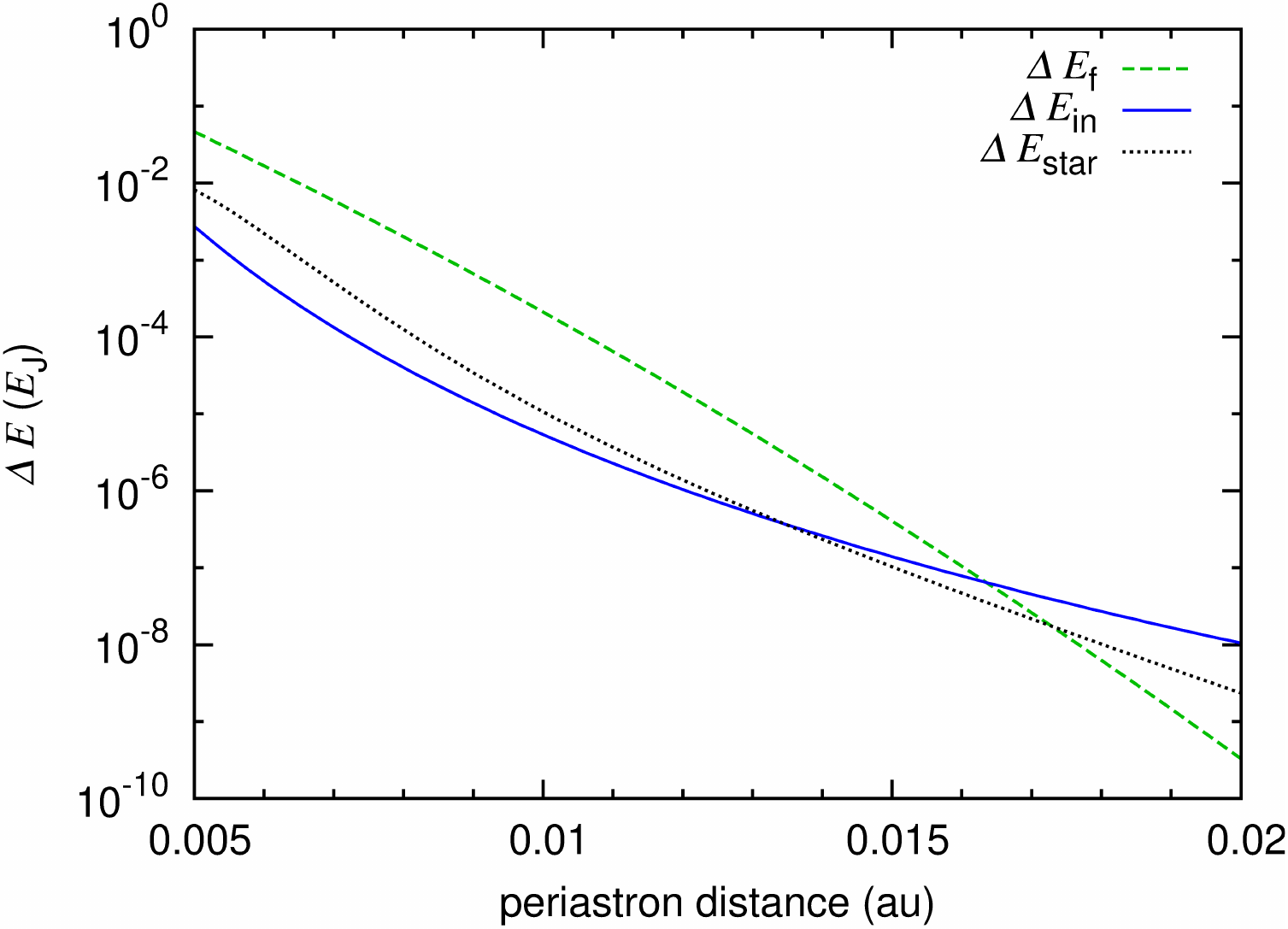}
\caption{\label{figtide} The energy of tidal transfer during one periastron passage.
A solar-mass and solar-radius star and a Jovian-mass and Jovian-radius planet are assumed. The energy transfer associated to the planetary fundamental mode $\Delta E_{\rm f}$ is given as a function of the periastron distance  by a green broken line, the planetary inertial mode $\Delta E_{\rm in}$ by a blue line and the energy associated to stellar tides $\Delta E_{\rm star}$ by a black dotted line. 
}
\end{figure}

We neglect the contribution from dynamical tides raised by the planet on the host star according to the indications of Figure~\ref{figtide}. In this  Figure  we compare the values of the energy transfer $\Delta E_{\rm f}$ and $\Delta E_{\rm in}$, in unit of $E_{\rm J}=Gm_{\rm J}^2R_{\rm J}^{-1}$, with tides associated with spherical harmonic index $\ell=2$ and $\ell=3$ in the star whose polytropic index is $n=3$  (hereinafter $\Delta E_{\rm star}$) as given by \cite{Lee1986}.
Here, $G$ is the gravitational constant, $M_{\rm J}$ is the mass of Jupiter and $R_{\rm J}$ its radius. To produce the curves in Figure~\ref{figtide} we assume a  planet with mass $m_p=1 M_{\rm J}$ and  physical radius $R_p=1 R_{\rm J}$. The stellar mass is set to 1 solar mass ($M_S=1 M_{\rm \odot}$) and its radius is 1 solar radius ($R_S=1 R_{\rm \odot}$). Following these choices of the parameter values,  $(M_S/m_p)(R_p/R_S)^3=1$. 
According to the predicted behavior of $\Delta E_{\rm f}$, $ \Delta E_{\rm in}$, and $\Delta E_{\rm star}$, we can guess that inside of $\sim$0.015 au planetary fundamental modes dominate the evolution of the system while beyond $\sim$0.015 au, planetary inertial modes dominate the orbital evolution. The energy exchange due to the stellar tides ($\Delta E_{\rm star}$) is always lower than that of the planetary tides ($\Delta E_{\rm f}$ and $\Delta  E_{\rm in}$) in both regimes. 

The trajectories of the planets are numerically integrated  using a 4th-order time-symmetric Hermite code \citep{kokubo98}.
Since at every periastron passage we need to update the energy due to the tidal interaction, we adopt a variable shared time--step. 
A typical time step is of the order of $2^{-9}r^{3/2}(GM)^{1/2}$, where $r$ is mutual distance and $M$ is the total mass of the pair of interacting bodies.

In modeling  P--P scattering events we consider 3 equal mass Jupiter planets ($m=M_J$) around a solar mass star. The force per unit mass acting on planet {\it i} 
is given by: 
\begin{equation}
{\bf F}_i= -\frac{G(M_*+m_i)}{r_i^3}{\bf r}_i
  +\sum_{j\neq i} Gm_{j}\left( \frac{{\bf r}_{j}-{\bf r}_i}{|{\bf r}_{j}-{\bf r}_i|^3} 
  - \frac{{\bf r}_{j}}{|{\bf r}_{j}|^3}\right)
+{\bf F}_{{\rm GR},i}.
\label{eq:basiceq}
\end{equation}
where the first term is the gravity from the star, the second term is the gravity from the other planets $j$, the third term is the indirect term due to the barycenter motion and the forth term is the general relativity contribution computed in the  post-Newtonian approximation with the correction terms given by \cite{kidder1995}.

When two planets approach each other at a distance smaller than the sum of their radii, we assume that a collision occurs and  merge the two bodies  into a single one. Its new velocity vector is computed from the conservation of momentum 
while its radius is derived adopting the same density of the other planets so that $m_p \propto R_p^3$. We then continue the simulation with the new bigger planet. Due to the largest radius of the merged body the tidal force becomes slightly stronger.  Planets which hit the star are removed from the simulation while those that are fully circularized are no longer evolved in time. 
%

\section{P--P scattering outcomes}

\begin{figure}[hpt]
\includegraphics[width=1.0 \hsize]{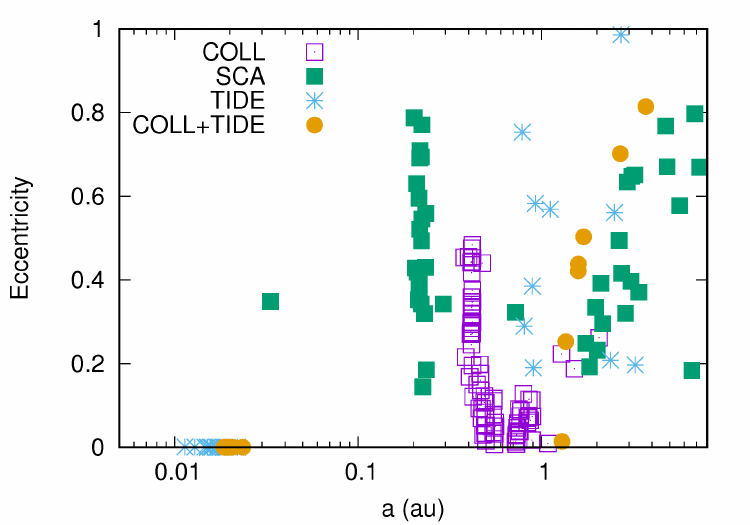}
\includegraphics[width=1.0 \hsize]{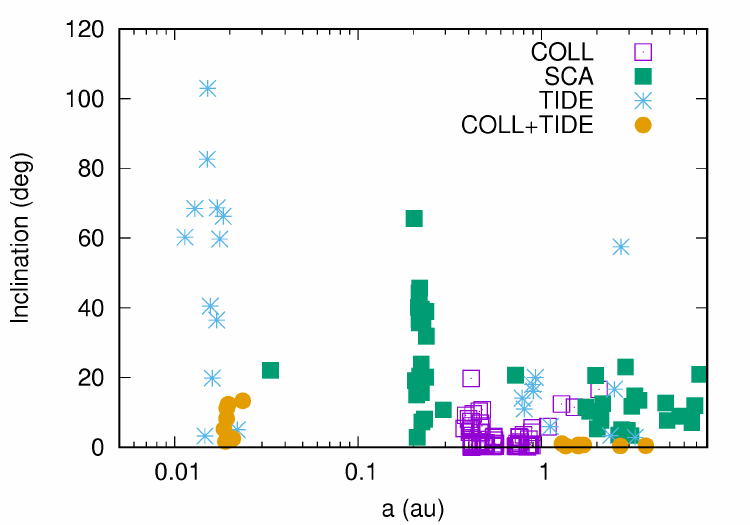}
\caption{\label{fig1} Eccentricity vs. semi--major axis (top plot) and inclination vs. semi--major axis  (bottom plot) of P--P scattering outcomes for three Jupiter --size planets started on circular orbits with  $a_1 = 1$ au, $a_2 = 1.5$ au, $a_3 = 2$ au. All other initial orbital elements are selected randomly, the orbital angles between [0,2$\pi$] and the inclination smaller than $1^o$. The time evolution of each system is 2 My.}
\end{figure}

At the end of the chaotic evolution of three Jupiter size planets,  we outline four main distinct possible outcomes:   
\begin{enumerate}
\item ejection of one planet out of the planetary system, two planets are left on eccentric orbits and the inner one has a semi--major axis determined by conservation of orbital energy. This is the classical pure N--body P--P scattering event
\item collision between two planets, both merger and scatterer are left on eccentric and inclined orbits or an additional collision occurs leaving a single eccentric planet in the system
\item ejection of one planet and fast tidal circularization of the inner one whose orbit shrinks enough to avoid further close encounters with its scatterer.
\item collision between two planets and tidal circularization to a close orbit of the inner one which may be either the more massive or the less massive of the two survivors. 
\end{enumerate}

Both events 3) and 4) may lead to hot Jupiters on orbits that can be significantly misaligned respect to the star equator.  

In Figure~\ref{fig1} we show the typical output of a numerical simulation with three equal mass Jupiter planets. They are initially set on orbits with semi--major axis of 1, 1.5 and 2 au, respectively, all circular and with a random initial inclination lower than $1^o$ while all the other orbital angles are randomly chosen between [0,2$\pi$] . The initial orbits are close enough to lead to instability on a short timescale and are a reasonable outcome either of convergent migration or of 'in situ' formation. The evolution of the system is followed for 2 Myr and, at the end, the different behaviors are grouped according to the previous classification. The classical scattering events bring the inner planet on an eccentric orbit around 0.2 au, according to the conservation of the orbital energy of the system. Two--planet collisions give rise to a second population of inner eccentric planets located close to 0.4 au. On average, these planets are less eccentric compared to those coming out of pure scattering events and their possible evolution is outlined and discussed in more detail in Figure~\ref{ecc_coll}. In a significant number of cases the inner planet decouples from the outer one and is circularized by tides to the pericenter of the scattered orbit,  around 0.01-0.03 au. The  inclination distribution shows that these last planets are often on highly misaligned orbits and one is even on   a retrograde orbit.

These results are in agreement with the findings of \cite{naga08} even if a larger number of collisional events are found due to the initial closer orbit of the inner planet \citep{petrovich2014}. However, in spite of the large number of collisions, the events 3) and 4) supply a significant number of misaligned Hot Jupiters while events 1) and 2) lead to eccentric and potentially misaligned planets  on close orbits even when a collision occurs.

The final location of close--in giant planets in our models reflects the strength of the tides that we include in our modeling, which play a very important role in the decay of planetary orbits. These are dynamical tides (e.g., \cite{lai97, ivanov2004, ivanov2007, ivanov2011}) and in our simulations we use a formulation given by \cite{ivanov2007} as described in Section\ref{numerical}. 
However, the impulse approximation used in the evaluation of dynamical tides becomes a poor approximation when the circularization proceeds (e.g., \cite{mardling1995a, mardling1995b} and the eccentricity becomes low. Equilibrium tides become then effective (e.g. \cite{beauge2012}  and references therein) and the tidal evolution may occur on a longer timescale. In short, at the beginning of the orbital evolution that leads to the formation of Hot/Warm Jupiters, dynamical tides are important in forcing the decay of the orbit. In the last part of the dynamical evolution when the eccentricity has become low, equilibrium tides are more important in determining the location where the planet stops. Unfortunately, at present it is not known when and how the two tides switch. When we change the magnitude of two tides, the final location of the planets can be adjusted 
\citep{beauge2012}. However, rather than introducing artificial effects, we continue to use dynamical tides in our simulations even for low eccentricities but we stop our simulations when the energy decreasing due to the tide at the pericenter overcomes the orbital energy leading to a clustering of tidally circularized planets  around 0.02 au. 
However, the final distribution of the inclination of the planets  does not depend on this choice and highly misaligned planets would be produced anyway.
Since the tidal evolution of planets with arbitrary inclinations is still not well known, we assume that planetary inclination is not  significantly changed during tidal evolution \citep{bar_ogil2009}. Thus, the planets maintain the inclination  they have when the circularization begins. 

If the orbits of the planets are initially  set closer to the star, a larger  number of collisions occurs at the expense of the pure scattering events where a planet is ejected from the system. In Table 1 we report the statistical distribution of the different outcomes of P--P scattering for increasingly smaller initial semi--major axis of the inner planet $a_1$. Below 0.5 au the number of pure scattering events drops quickly to zero and there is a consistent reduction of tidal 
circularized planets in this kind of events. On the other hand, there is an increase of events where, after a collision,   the merged planet, or its lighter companion,  is progressively circularized by tides. 

\begin{table}[hpt]
\caption{Statistics of P-P scattering events as a function of the distance from the star}             
\label{table:1}      
\centering                          
\begin{tabular}{c c c c c}        
\hline\hline                 
$a_1 (au) $ & Scatt & Coll & Tide & Tide+Coll \\    
\hline                        
   5 & 67 & 12  & 22  &  0 \\      
   1 & 30 & 58   & 13   & 0 \\
   0.5 & 23  & 60  & 11   & 7 \\
   0.05 & 0 & 95 & 1 & 4\\
   0.03 & 0 & 83 & 1 &15 \\

\hline                                   
\end{tabular}
\end{table}

In Section 5 we will discuss the orbital distribution of the planets for different initial values of $a_1$. 

\section{Effects of general relativity on the secular evolution of two planets}

The main effect of general relativity ( hereinafter GR) is to induce the precession of the periapsis of 
a planet orbit according to the following equation (in the PN approximation): 

\begin{equation}
\frac {d \omega} {dt }  = \frac {3 G^3 (M_S + m_p)^{\frac {3}{2}} }{ a_p^{\frac{5}{2}} c^2 (1-e_p^2)},
\label{eq:C21}
\end{equation}

\citep{misner73}. The precession rate is faster for planets close to the star  and on  eccentric orbits due to the $(1-e_p^2)$ term at the denominator. This implies that if the outcome of a planet--planet scattering event injects a planet in an inside and eccentric orbit, Kozai--Lidov oscillations  of eccentricity due to an external  perturbing planet are gradually quenched  \citep{naoz2016}.  It is then expected that the number of tidal circularization is somewhat reduced, but not 
significantly, since the timescale of the Kozai cycle induced by an outer planetary companion may be
fast \citep{naga11} in particular in the close configurations we are considering in this paper. As a consequence, we do not expect significant changes due to the interaction between the Kozai--Lidov effect and relativity.

\begin{figure}[hpt]
\includegraphics[width=1.0 \hsize]{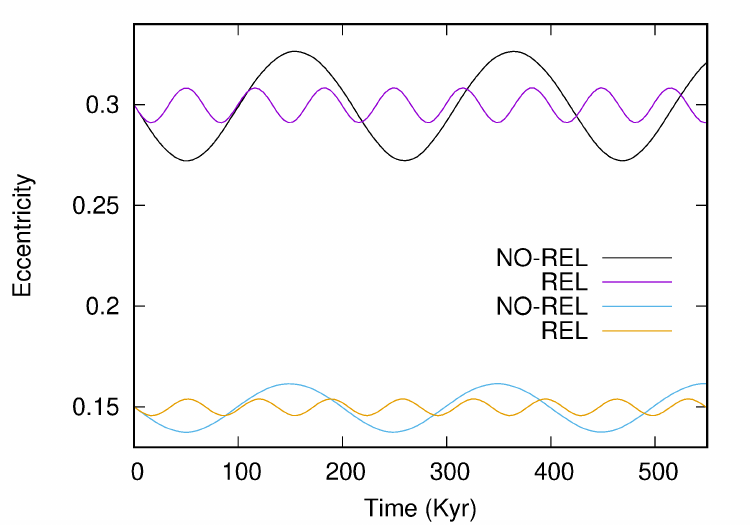}
\caption{\label{fig_sec0} Secular evolution of the eccentricity of the inner planet of a pair whose orbits have been integrated  with and without the GR term. The semi--major axis of the inner planet is  $a_1 = 0.1$ au while its eccentricity is set to $e_1 = 0.15 $ and $e_1 = 0.3$  in the two different cases. The outer planet has $a_2 = 1.5$ au and its eccentricity is  $e_2 = 0.1$. The mutual inclination is initially set to $3^o$. Tidal forces are not included in these models.}
\end{figure}

\begin{figure}[hpt]
\includegraphics[width=1.0 \hsize]{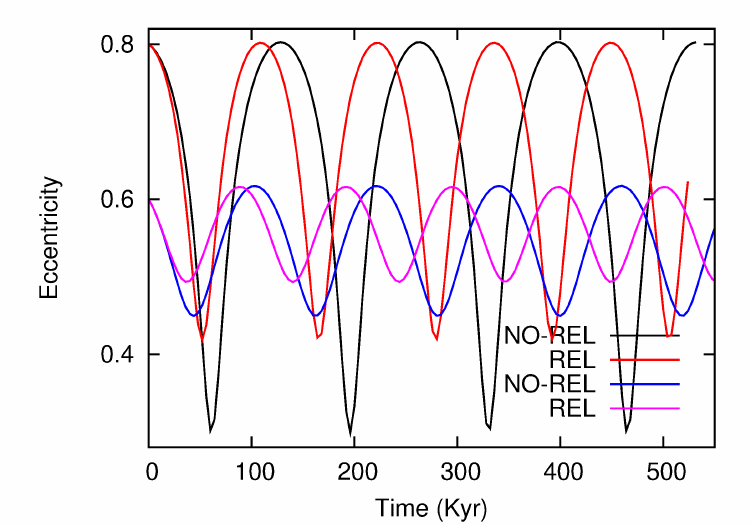}
\caption{\label{fig_sec1} 
Same as in Figure~\ref{fig_sec0}  but the semi--major axis of the inner planet has been shifted outwards to $a_1 = 0.2$ au to show that GR is relevant even further out. The initial eccentricities of the planets are $e_1 = 0.6$ and $e_1 = 0.8$ while that of the outer planet is fixed to $e_2 = 0.1$.}
\end{figure}

\begin{figure}[hpt]
\includegraphics[width=1.0 \hsize]{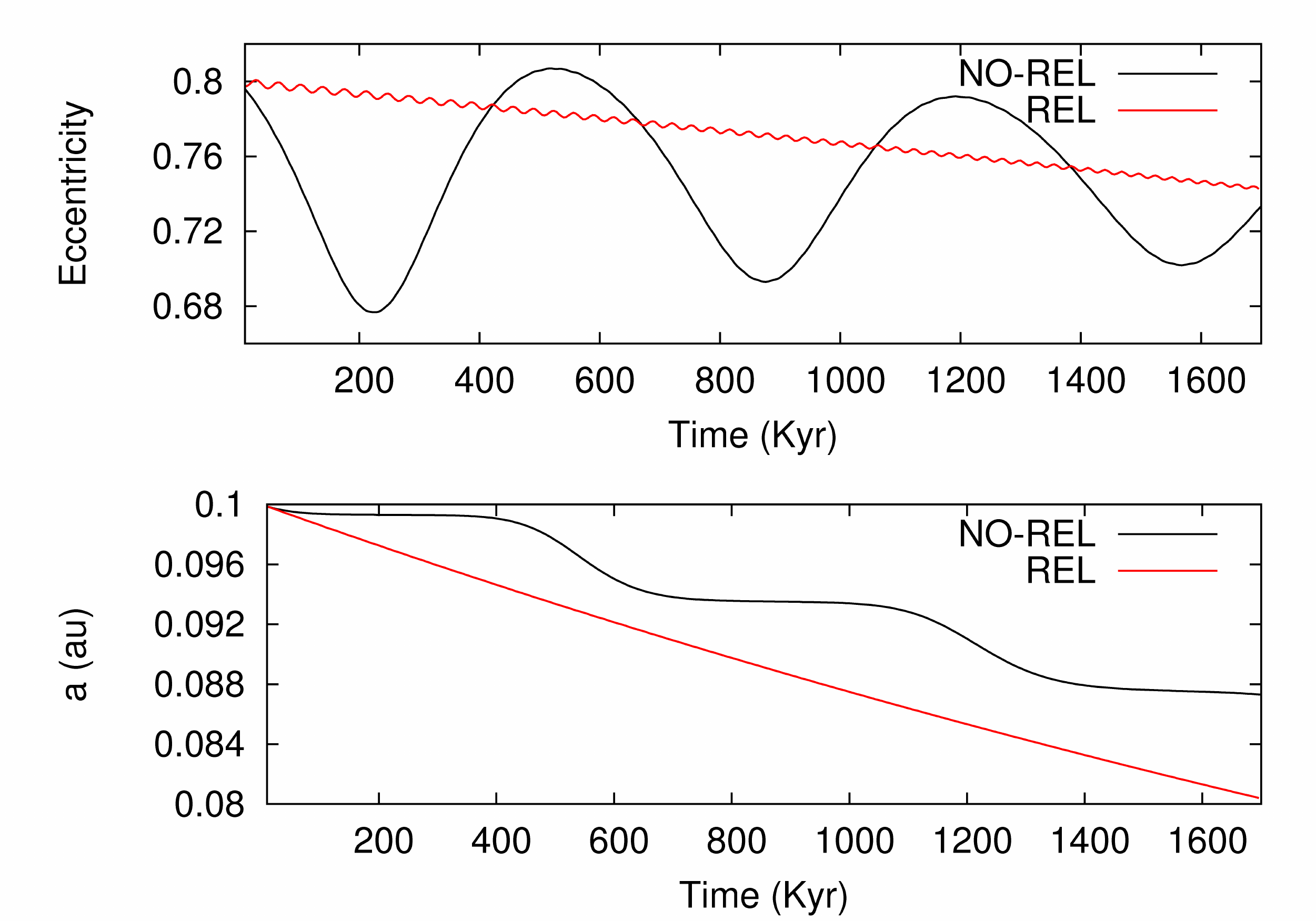}
\caption{\label{fig_sec2} As in Figure~\ref{fig_sec1} but the semi-major axis of the inner planet is set initially to 0.1 au and tidal forces are included. In the top panel the eccentricity 
of the inner planet is plotted with and without GR and both show the effects of tidal damping. }
\end{figure}

On the other hand,  if the architecture of the planetary system does not involve significantly misaligned planets that can lead to a  Kozai--Lidov evolution, GR  is relevant since it can significantly affect the 
the classical secular oscillations of  eccentricity and perihelion longitude. In  Figure~\ref{fig_sec0}  we illustrate the evolution of a planet close to the star with semi--major axis $a_1 = 0.1$ au perturbed by an outer planet with $a_2 = 1.5$ au. The mutual inclination is $3^o$ deg and  the initial eccentricity is set to 0.15 and 0.3 for the inner planet and 0.1 for the outer one. We neglect in these models the contribution from tidal forces.  While the average forced eccentricity remains the same in the case with and without GR, the proper oscillations are significantly shrunk when the GR effects are accounted for. This can be understood, as a first approximation,  in the framework of the second order secular theory of Laplace--Lagrange \citep{murray-dermottSS}. The equation for the non--singular variables $h_i = e_i sin(\varpi_i) $ after the inclusion of the GR term becomes:

\begin{equation}
\dot h_i = k_1  \left ( A_{i1} +  \frac {3 G^3 (M_S + m_p)^{\frac {3}{2}} }{ a_p^{\frac{5}{2}} c^2 (1-e_p^2)} \right ) +  k_2 A_{i2} ,
\end{equation}

where $k_i = e_i sin(\varpi_i) $ while the matrices $A_{i,j}$ are given in \citep{murray-dermottSS} and are only functions of the semi--major axes and masses of the two planets. While for the first planet, and then for $h_1$,  the additional term is relevant,  for the second planet ($h_2$), the GR term can be safely neglected since the body is too far from the star for GR to be significant. The increase in the  first diagonal term of the matrix $A_{ii} $ leads, in addition to a larger difference between the two main frequencies of the system and then to a faster oscillation frequency of the eccentricity,  to an increase in the modulus of the term $e_{11}$ of the first eigenvector associated to the secular evolution of the system.  When we compare the evolution of two planets with and without GR but starting with the same initial conditions, we expect that an increase in $|e_{11}|$ leads to a consequent decrease of $|e_{12}|$, the first element of the second eigenvector. However, to comply with the fixed initial conditions, the term $e_{11}^2 + e_{12}^2$ must remain constant (both models start with $\varpi_1 - \varpi_2 = \pi/2$) and, as a consequence, an increase in  $|e_{11}|$ leads to a decrease in 
$|e_{11} e_{12}|$ which is the half--width of the secular oscillation. This is confirmed by the behavior shown in Figure~\ref{fig_sec0}  where the amplitude of the secular oscillations of the case with GR is reduced compared to those without GR and the frequency is higher. 

If we further increase the eccentricity of the inner planet, as in  Figure~\ref{fig_sec1}, the situation becomes more complex since the second order secular theory does not properly work at high eccentricities and higher order expansions of the perturbative potential are required as suggested in \cite{henrard2005}. In this context, the interpretation of the numerical results on the basis of an  analytical theory is more challenging and we prefer to show the outcome of the numerical integrations and guess a general trend.  We again focus on two different models with different initial eccentricity and we also change the initial semi--major axis of the inner planet to show that the effects of GR on the secular evolution are relevant even farther from the star if the eccentricity is high.  In Figure~\ref{fig_sec1} we show the case of a pair of planets where the inner one has a semi--major axis of 0.2 au and an initial eccentricity of 0.6 and 0.8 in the two different cases, while the outer one has  $a_2 = 1.5$ au and $e_2=0.1$. As in the low eccentricity cases illustrated in Figure~\ref{fig_sec0}, the proper oscillation is reduced in amplitude but, differently from the low eccentricity cases, the forced term is larger. As a result, the peak eccentricity is approximately equal in the two cases (with and without GR) but the mean eccentricity in the model with GR is higher because of the smaller oscillation amplitude.  This has significant implications for the interpretation of a statistical sample of giant exoplanets and for the high--eccentricity migration mechanism \citep{petrovich_secular_2016}.  If we analyse a set of exoplanetary systems with two planets where one is close to the star,  a prediction of their expected mean eccentricity based on a secular model without GR would underestimate its mean value. By inspecting Figure~\ref{fig_sec1} it can be inferred  that GR, by increasing the bottom value of the secular oscillation, gives a higher mean eccentricity. 

The growth in the average eccentricity of the inner planet   significantly  influences  also its tidal evolution. In  Figure~\ref{fig_sec2} we illustrate the evolution of a system where  the semi--major axis of the inner planet is  $a_1 = 0.1$ au and the initial eccentricity is set to 0.8, the same value of the high eccentricity case in Figure~\ref{fig_sec1}. In this model we also include the tidal force on the planets to test the combined effects of tides and  GR.   The evolution of the eccentricity when the GR effects are included shows high frequency tiny oscillations around an almost linearly decreasing trend due to tidal damping. This fast oscillations indicate that the evolution of 
$\varpi_1$ is dominated by the GR term which also forces the eccentricity to remain constantly high. Without the GR term, the secular evolution shows wide oscillations also damped by the tidal interaction but on average the planet eccentricity is lower respect to the GR case. This different behavior of the eccentricity affects also the semi--major axis evolution. With GR, the  eccentricity is slowly damped but the permanence of the planet in a highly eccentric orbit favors a faster and smoother inward migration of the inner planet (bottom panel of Figure~\ref{fig_sec2}) compared to the case without GR. 
The so called secular high--eccentricity migration \citep{petrovich_secular_2016} is then enhanced and GR  leads to a faster inward drift of the inner planet compared to the secularly oscillating case without relativity. This behavior favors the evolution of eccentric giant planets into Hot/Warm Jupiters on circularized orbits.  

\section{P--P scattering close to the star}
\label{PPclose}

To gain a better insight on the outcomes of P--P scattering events occurring close to the star and on how they compare to the observed orbital distributions of the known exoplanets, we have performed three different sets of simulations where the initial semi--major axes of the inner planet $a_1$ is randomly selected between 0.03-0.05 au, 0.05-0.1 au, and 0.2-1.0 au, respectively.  The other two planets are started on orbits separated by $3  R_H$ where $R_H$ is the Hill's sphere. The results are shown in  Figure~\ref{fig_models} for all cases and they are compared to the observed distributions of exoplanets more massive than 0.5 $M_J$ (\cite{enciclo}, http://exoplanet.eu/)  added in the bottom panels.  In the 
semi--major aixs vs. eccentricity distribution of observed exoplanets (left--bottom panel of Figure~\ref{fig_models} there are two noticeable over--densities,  one close to 0.05 au and the second just beyond 1 au. These peaks may tell us something about the dynamical evolution of giant planets or they may simply be due to an observational bias since most of the very  close planets are detected by transits while the outer ones have been mostly revealed by radial velocities. The pileup in the proximity of 1 au suggested by some authors as due to inhomogeneities in the protoplanetary disks related to  dead zones, ice lines, heat transitions  \citep{hasegawa2012} or photoevaporation \citep{matsuyama2003,alexander2009,alexander2012,ercolano2015}, has been recently re-evaluated and questioned by \cite{wise2018,wise2018-2}.  Additional detections are needed to have the final say on the (a,e) distribution of giant planets close to their star and on the origin of it.  Very few data are presently available on the planet orbital misalignment  respect to the rotation axis of the star as it can be seen in the bottom right panel of Figure~\ref{fig_models} where all the available data, obtained thanks to the Rossiter-McLaughlin effect, are reported. 

If we combine the outcome of our modeling where $a_1 = 0.03-0.05$ au (top left panel) and $a_1 = 0.05-0.1$ au (second left panel), the surviving inner planets well reproduce the first observed peak in the semi--major axis versus eccentricity distribution (lowest left panel). High eccentricities are not achieved in the numerical models for semi--major axes smaller than 0.1 au and this is partly due to the merging of two planets after a collision and to tidal effects. In these models  the inner planet may be either the more massive one, outcome of a single or even double merging event, or the lighter one not involved in a collision.  According to observations, close exoplanets cover a wide range in  mass and the majority is encompassed within 4 masses of Jupiter and this is compatible with the prediction that these close planets are the outcome of a collision between two Jupiter size planets. The smaller ones might be the result of an impact between less massive planets possibly driven in close orbits by either type I migration or trapped and transported by  disk inhomogeneities or formed in situ. 

\begin{figure}[hpt]
  \centering  
    \includegraphics[width=1.0\linewidth]{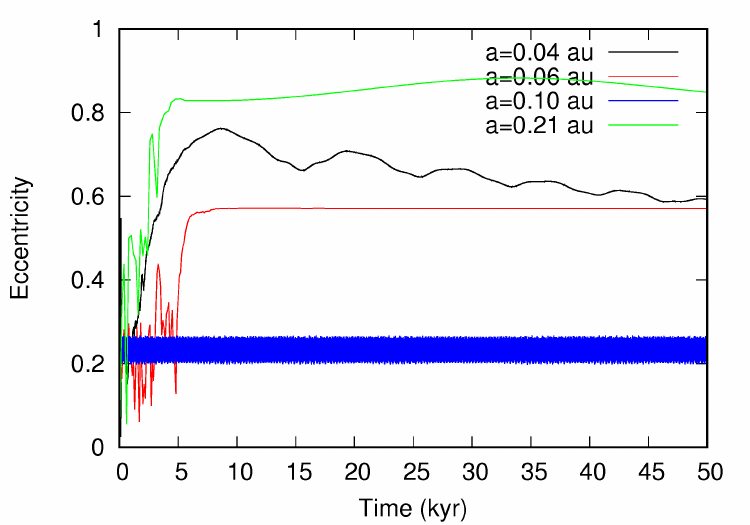}
\caption{\label{ecc_coll} Eccentricity evolution of the planet ending up in the inner orbit after a collisional event involving two of the three initial planets.  Different starting values of the planet semi--major axes are adopted. The legenda within the figure gives the final semi--major axis of the inner planet when a dynamically stable configuration is reached.  Different dynamical final states are illustrated.  The blue line shows a regular secular evolution of the two surviving planets after the collision. The black line shows a sharp grow of the eccentricity during the chaotic evolution followed by a slow tidal damping of the inner planet,  the less massive of the two. The red line shows a case where two collisions occur and a single massive planet is left on an eccentric orbit. The green line shows the case in which the merged planet ends up into a highly eccentric orbit and its evolution is decoupled from the outer one.}
\end{figure}

\begin{figure*}
  \centering
  \begin{subfigure}[b]{0.45\textwidth}
    \centering
    \includegraphics[width=1.0\linewidth]{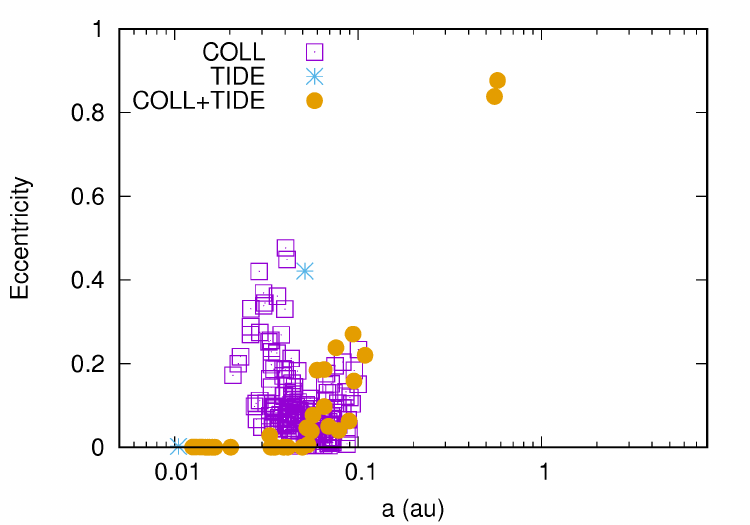}
  \end{subfigure}%
  \quad
  \begin{subfigure}[b]{0.45\textwidth}
    \centering
    \includegraphics[width=1.0\linewidth]{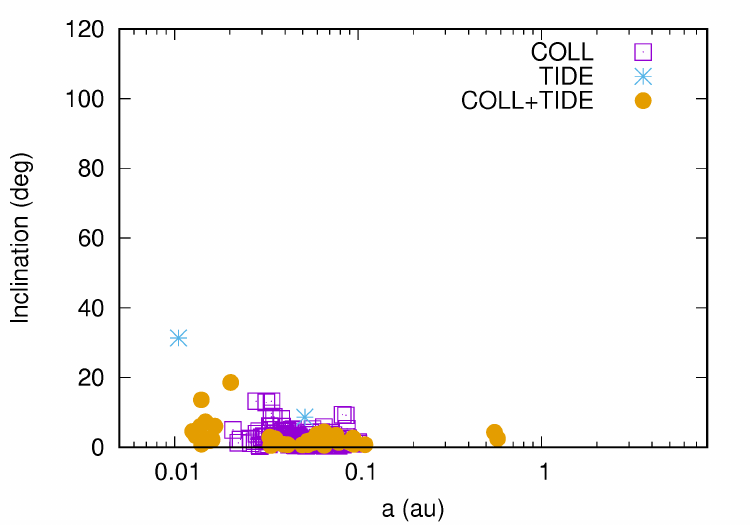}
  \end{subfigure}
   \centering
  \begin{subfigure}[b]{0.45\textwidth}
    \centering
    \includegraphics[width=1.0\linewidth]{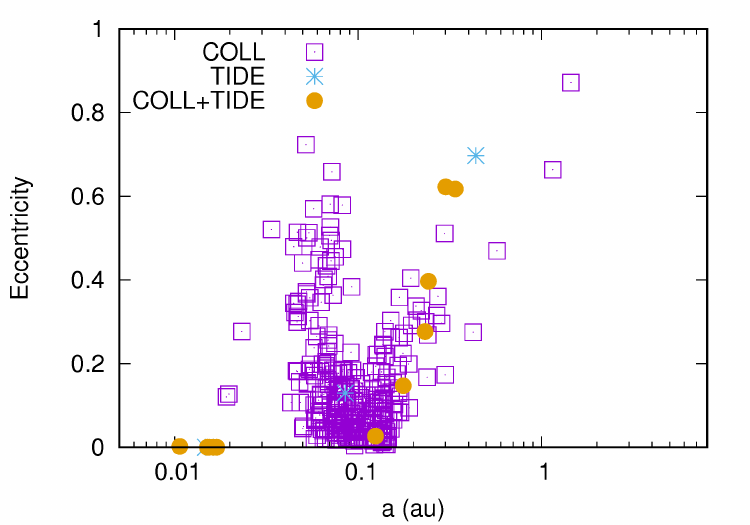}
  \end{subfigure}%
  \quad
  \begin{subfigure}[b]{0.45\textwidth}
    \centering
    \includegraphics[width=1.0\linewidth]{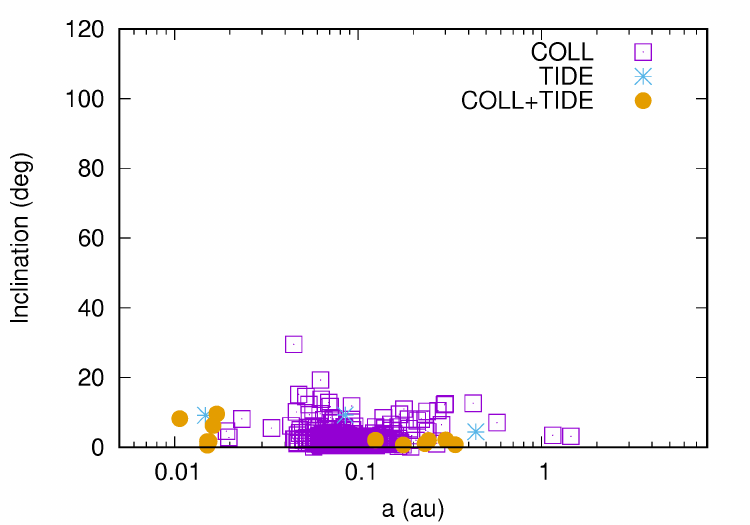}
  \end{subfigure} 
  \centering
  \begin{subfigure}[b]{0.45\textwidth}
    \centering
    \includegraphics[width=1.0\linewidth]{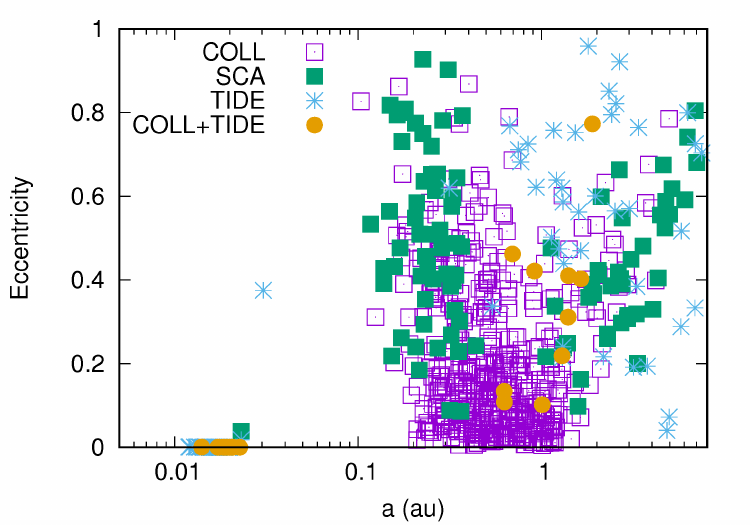}
  \end{subfigure}%
  \quad
  \begin{subfigure}[b]{0.45\textwidth}
    \centering
    \includegraphics[width=1.0\linewidth]{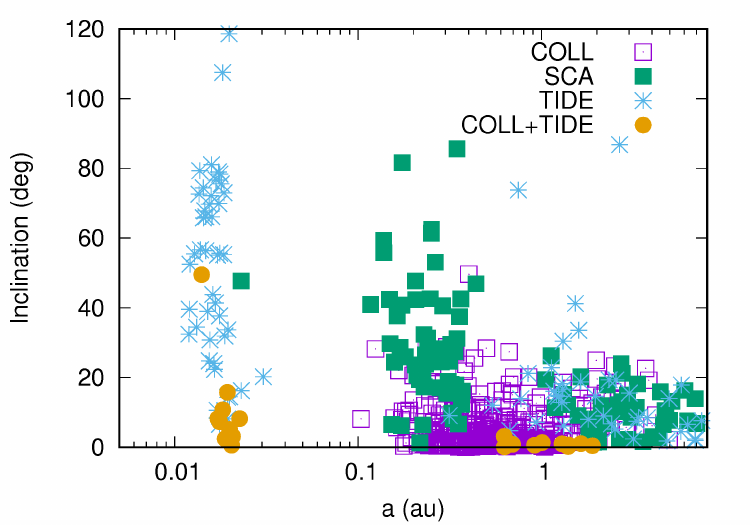}
  \end{subfigure}
  \centering
  \begin{subfigure}[b]{0.45\textwidth}
    \centering
    \includegraphics[width=1.0\linewidth]{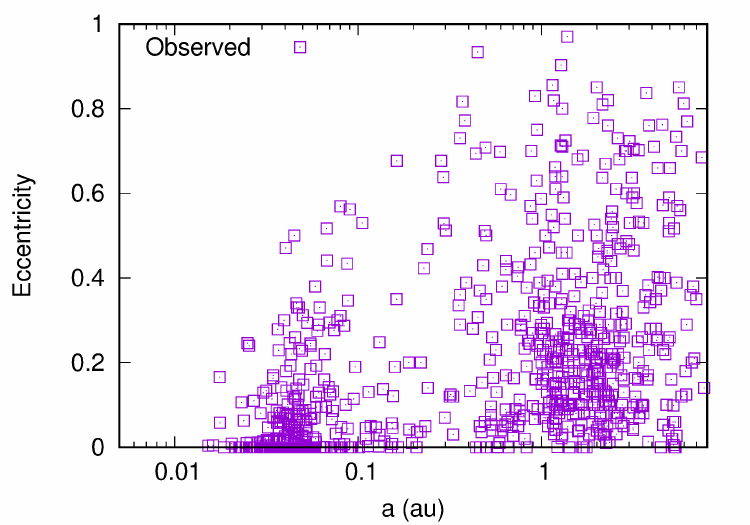}
  \end{subfigure}%
  \quad
  \begin{subfigure}[b]{0.45\textwidth}
    \centering
    \includegraphics[width=1.0\linewidth]{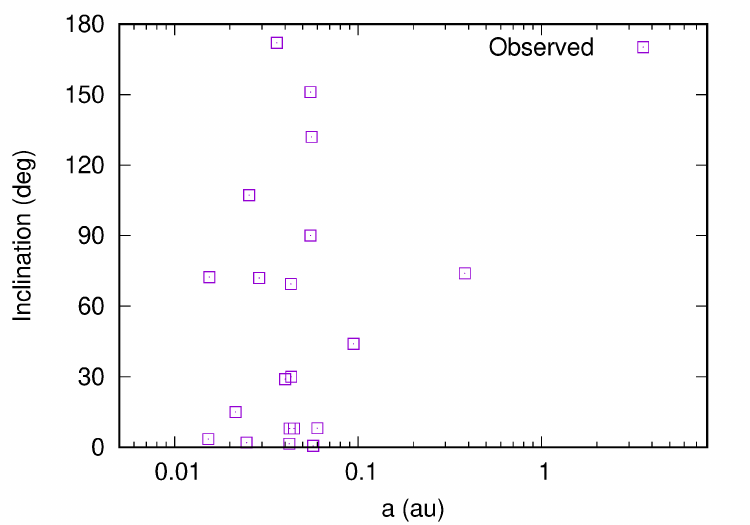}
  \end{subfigure}
  \caption{Comparison among model orbital distributions for different initial values of the inner planet semi--major axis $a_1$  and the distribution of presently known exoplanets \citep{enciclo}. In the top panels $a_1$ is randomly selected within the interval 0.03-0.05 au, in the second row panels within  0.05-0.1 au, in third row panels within [0.2-1]au. In the bottom panels we show the observed orbital distribution.  The inclination respect to the star equator ($\lambda$ parameter) is known only in a limited number of cases. }
  \label{fig_models}
\end{figure*}

To interpret the outcome of the simulations shown in the two upper panels of Figure~\ref{fig_models}, it is important to investigate the dynamical evolution of the planets after a collision between two of them. The planetary system, after a collision and merging, may  follow different evolutionary paths
(see Figure~\ref{ecc_coll}) . The two surviving planets can end up in a stable secular configuration where they evolve in time according to the standard linear secular theory  (blue line in Figure~\ref{ecc_coll}).  In alternative, after the collision, the surviving planets are still in a chaotic state where the orbits cross each other. During this instability the eccentricity of both planets is further excited until either they collide, leaving a single massive planets in the system on an eccentric orbit (red line in Figure~\ref{ecc_coll}), or the inner planet is affected by tides and its orbit decouples from that of the outer planet (green and black line of Figure~\ref{ecc_coll}).  In this last case, the inner planet can be either the more massive or the lighter one. After the dynamical decoupling, the system will follow two different paths depending on the distance of the planet from the star. If the planet is in a very close orbit it may be  affected by tides, as in the case illustrated by the black line in Figure~\ref{ecc_coll}, and be circularized to the pericenter, or it may remain on a highly eccentric orbit if it is not close enough to the star to be significantly affected by tides. In most of these configurations, the final outcome is a system with two planets close to the star and on eccentric orbits. Among the different final architectures of the planets after the chaotic phase and a collision, there are cases where the inner planet is circularized by tides while the outer one is close to the inner planet but dynamically decoupled from it. This last planet is a good candidate to be detected by TTV (Transit Time Variations) on the  orbit of the inner one  due to the strong mutual secular perturbations.  

An aspect that is not well reproduced by the models where the scattering/collision occurs close to the star (two upper panels of  Figure~\ref{fig_models} is the inclination distribution.  All simulations with the inner planet starting within 
0.1 au, the spin/orbit misalignment is typically lower than $20^o$.  These low inclinations do not fully comply with observations since exoplanets found on  close orbits may  also be significantly misaligned respect to the spin axis of the star (lowest right panel of Figure~\ref{fig_models}).
Ad example, the hot Saturn KELT-6b \citep{damasso2015} is on a misaligned orbit with the projected spin-orbit angle $\lambda$, measured thanks to the Rossiter-McLaughlin effect, of about $36^o$. 
As a consequence,  the inner peak in the orbital distribution of observed exoplanets can be well reproduced by P--P scattering events with the exception of misaligned Hot/Warm Jupiters. 

This apparent conflict between models and observations can be solved in two ways. One possibility is that  the planets are driven into close orbits on already inclined orbits, but we have already argued that this cannot happen in presence of the gaseous disk which quickly damps the inclination. In alternative, our simulations show that the problem can be  solved when we consider P--P scattering events taking place farther out from the star. In the third row of Figure~\ref{fig_models} we draw the orbital distribution of surviving planets from a population where the inner planet had its semi--major axis randomly distributed in between 0.2-1.0 au.  The semi--major axis vs. eccentricity distribution show a density peak of eccentric planets  around 1 au with eccentricity values larger than 0.9 coming mostly from scattering events. This peak  may conceivably be correlated with the observed one, even if it is related to the particular choice of starting conditions.   We may easily deduce that if we slightly increase the value of $a_1$,  we can move the peak outward obtaining an even better match to the 1 au pileup, although it may be a bias artifact.  The important aspect of the P--P scattering events occurring within $a_1=0.2-1.0$ au is that they supply a significant number of circularized planets on misaligned orbits. In the right panel of the third row in Figure~\ref{fig_models} we note a significant number of these planets  that fill the region left empty by the previous models where the inner planet was initially closer to the star. The misaligned Jupiters are, in the majority of cases, the output of scattering events without collisions and, in a limited number of cases, even retrograde orbits are obtained. 

When comparing  modeling outcomes with the real distributions of exoplanets we have to take into account that the simulations show the final distribution after the same fixed time-span of the numerical integration while the observed semi--major axis, eccentricity and inclination distributions group together systems with different evolutionary ages. This is a common problem in this type of modeling and it may lead to an overestimate or underestimate of the number of circularized systems. In the observed systems some planets had enough time to be further circularized compared to the simulations. However, in our numerical model dynamical tides tend to cause a fast evolution  when the eccentricity is low and static tides should play a major role. As a consequence, the two effects might somehow balance and make our results more robust. 




\section{Dependence on the radius of the planet}

\begin{figure}[hpt]
  \centering  
    \includegraphics[width=1.0\linewidth]{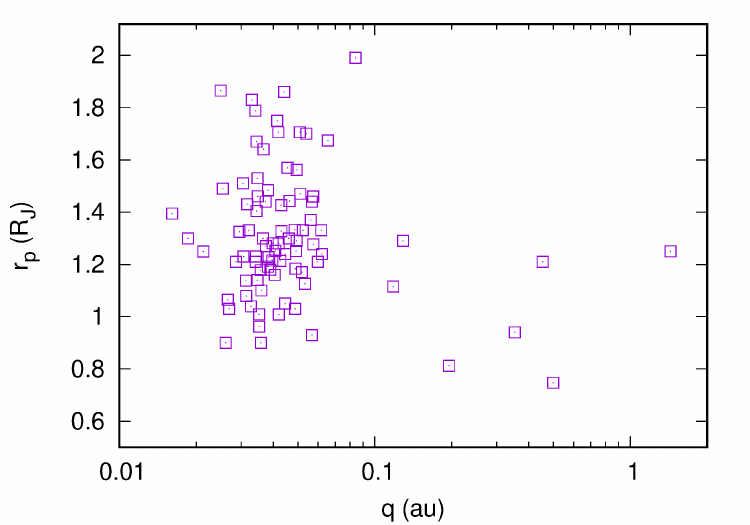}
\caption{\label{rm} Radii of all exoplanets known to date ($R_J$ is Jupiter's radius) with mass in between 0.8 and 1.2 $M_J$ ($M_J$ is Jupiter's mass) as a function of the pericenter distance. }
\end{figure}

 So far, in our simulations we have assumed for all planets a radius comparable to that of Jupiter. However,  the mass--radius relation may change depending on the orbital distance, stellar luminosity, age and composition of the planet. In particular, young planets are expected to undergo subsequent cooling and contraction so that their initial radius is inflated compared to that of mature planets. According to the models of \cite{fortney2007,mordasini2012}, where state of art equations of state and heavy element enrichment have been adopted,  young inflated Jupiter--like planets may reach radii as large as 1.5--1.6 $R_J$ where $R_J$ is Jupiter's radius. {\bf Besides model predictions, we can also look at the presently observed population of transiting exoplanets with known mass and radius. In Figure~\ref{rm} we plot the radius $r_p$ of all observed exoplanets to date in the mass range $0.8 M_j < m_p < 1.2 M_J$ 
(\cite{enciclo}, http://exoplanet.eu/) 
as a function of their pericenter distance from the star.  Most data concerns planets close to their host star since the radius is derived from transits and only a few values are known far from the star. If we limit the sample to only those planets with pericenter within 0.1 au and exclude the pulsar planet PSR 1719-14 b, the average value of  $r_p$ is 1.33 $R_J$ while the median is slightly smaller and equal to 1.28 $R_J$ with the two extremes being Kepler-435  with the largest radius (2.10  $R_J$) and  WASP-129 b  being the smallest one (0.93 $R_J$).}
In this sample of real planets it is however difficult to disentangle the different effects that contribute to the radius distribution due to the concomitant presence of planets with different ages and different heavy element content and core size. In Figure~\ref{rm} we expect to look at planets which are inflated either because of stellar irradiation or because they are young and not yet cooled to their final temperature. In addition, the different composition influences both these effects introducing an additional spread in the density values for a given planet mass. While the age may be estimated from that of the star, the composition at present is almost impossible to be determined. 

Since in our dynamical simulations both the collision probability and the tidal dissipation depend on the radius of the planets, we performed an  additional set of simulations with a larger radius for the planets (i.e. lower density). Instead of using an average value derived from observations (1.33 $R_J$), we prefer to deal with a more extreme case and we adopt for all three planets a radius of 1.5 $R_J$. This choice allows us to better explore the effect of an inflated radius on the P--P scattering dynamics and it also covers the majority of observed planets which, according to  Figure~\ref{rm}, mostly populate the region below this value. It could be argued that this value is indeed appropriate for young planets evolving close to the star. On the other side,  planet--planet scattering events do not necessarily occur when the planets are young since the onset of instability and chaotic behavior can occur even after some billions of years from the formation of the planets, depending on their mutual distance at the time of the disk dissipation \cite{weiden-marza96, chambers96}.  In the future, when the mass--radius relationship will be better defined as a function of the planet age and, possibly, composition, one may imagine to build up a numerical model for P--P scattering in which the radius changes with time. However, there is an intrinsic limit for this kind of modeling set by the integration time. If the planets are close to the star, a numerical simulation can cover time intervals of the order of only some  Myr with a reasonable CPU load. Beyond that, the time required to complete the runs becomes prohibitively long, preventing any chance of covering a significant range of evolution of the radius of the planets. 

\begin{figure*}
  \centering
  \begin{subfigure}[b]{0.45\textwidth}
    \centering
    \includegraphics[width=1.0\linewidth]{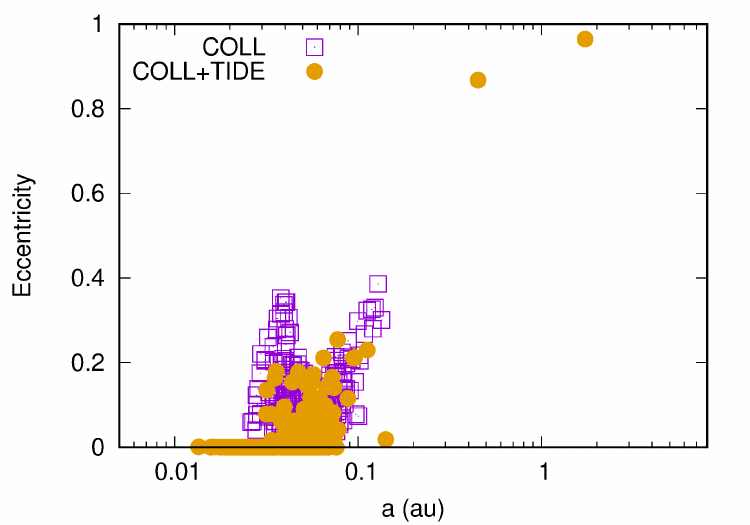}
  \end{subfigure}%
  \quad
  \begin{subfigure}[b]{0.45\textwidth}
    \centering
    \includegraphics[width=1.0\linewidth]{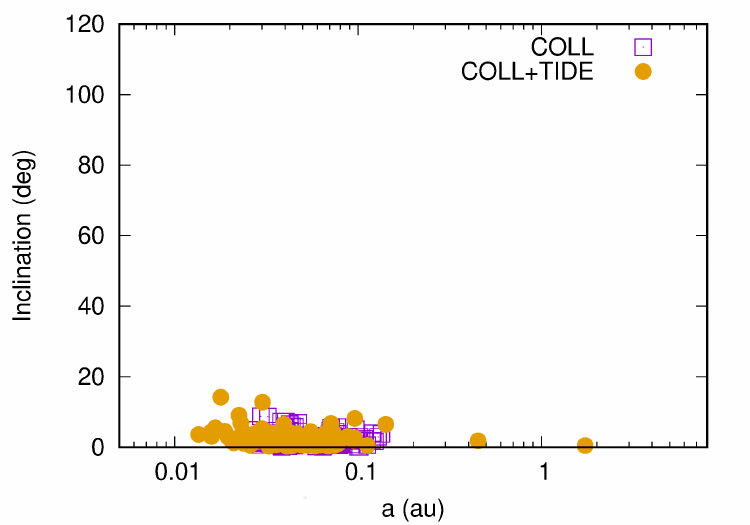}
  \end{subfigure}
   \centering
  \begin{subfigure}[b]{0.45\textwidth}
    \centering
    \includegraphics[width=1.0\linewidth]{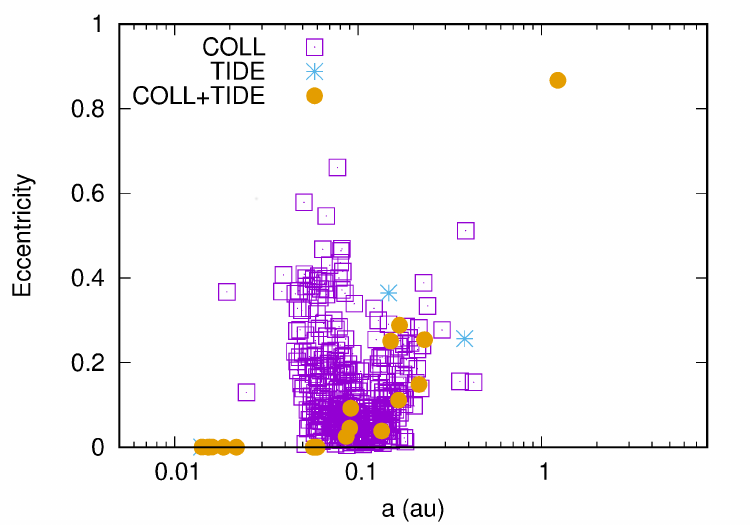}
  \end{subfigure}%
  \quad
  \begin{subfigure}[b]{0.45\textwidth}
    \centering
    \includegraphics[width=1.0\linewidth]{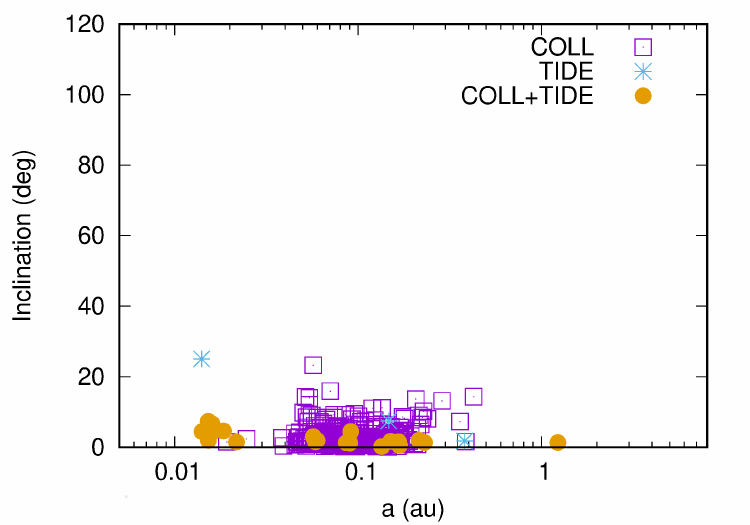}
  \end{subfigure} 
  \centering
  \begin{subfigure}[b]{0.45\textwidth}
    \centering
    \includegraphics[width=1.0\linewidth]{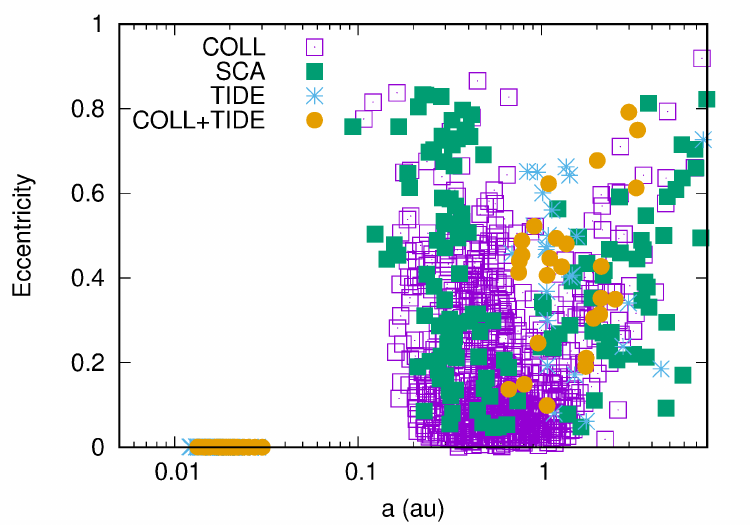}
  \end{subfigure}%
  \quad
  \begin{subfigure}[b]{0.45\textwidth}
    \centering
    \includegraphics[width=1.0\linewidth]{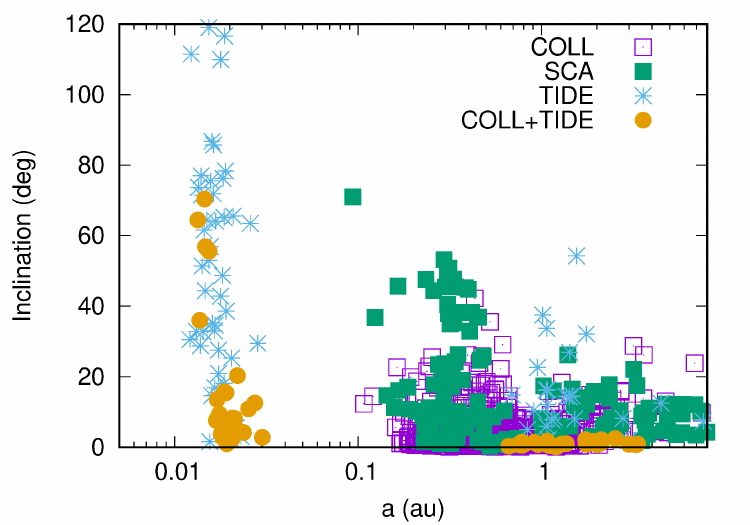}
  \end{subfigure}

  \caption{Same as Figure~\ref{fig_models} but with the radius of the planets increased to 1.5 $R_J$. }
  \label{fig_models_rad}
\end{figure*}

In Figure~\ref{fig_models_rad} the outcomes of the simulations with $r_p = 1.5 R_J$ are shown and can be compared to the cases illustrated in Figure~\ref{fig_models}. In the top two panels showing the P--P scattering events with the inner planet very close to the star ($a_1 = 0.03-0.05$ au) it is noteworthy that the number of planets with circularized orbits  is significantly increased. In the standard case with 
$r_p = 1.0 R_J$ the tidally circularized planets (after a collision) were 16\% while this percentage grows to 46\% in the case with $r_p = 1.5 R_J$.  This is an indication of the increased strength of the tide for inflated planets. However,  the overall final eccentricity distribution still shows eccentric bodies, in particular the companions of those scattered inwards. As a consequence, even for expanded planets we expect to find a significant number of eccentric ones close to their star and possibly cohabiting with an outer close--by planet causing observable TTVs.  The final inclination distribution does not show interesting differences in the two cases and only mild values of inclination are achieved.

When the initial semi--major axis of the inner planet is chosen in the range  $a_1 = 0.05-0.1 $ au, no  significant  differences are observed in the occurrence of the  different outcomes of the P--P scattering events compared to the case with  $r_p = 1 R_J$. There is instead a decrease in the number of highly eccentric planets with $e_p > 0.5$ when a larger radius is adopted and this is an additional indication of a stronger tidal circularization during the P--P evolution. Even in this case the final inclination distribution does not differ in the two scenarios. In the third row of  Figure~\ref{fig_models_rad} where the semi--major axis of the inner planet is moved further outwards  ($a_1 = 0.2-1.0$ au),  the  effects of the inflated radius is almost negligible  and the outcomes of the simulations are consistent with the case with $r_p= 1 R_J$.  The fraction of scattering events without collisions is approximately 15\% when $r_p = R_J$ and 13\% for $r_p = 1.5 R_J$. This limited reduction in the scattering probability suggests that in both models (at equilibrium or inflated planets) the scattering cases occur under similar conditions where a quick increase in inclination favors close encounters rather than collisions. We observe an increase in the number of tidally circularized planets after a collision when $r_p = 1.5 R_J$, an additional indication of the enhanced tidal force for inflated planets. The farther out we get from the star, the less inflated are expected to be the planets so we focus only on cases within 1 au where, also for inflated planets, we find a good 
agreement with the observed orbital distribution of exoplanets (Figure~\ref{fig_models}, bottom panels).

\section{Discussion and conclusions}

We have explored in this paper the dynamical outcome of P--P scattering events occurring when the planets are driven (or born) close to their parent star. The goal is to check if these events can explain the high eccentricities of a significant fraction of close--in Jupiter--size planets and the large misalignment respect to the planet equator measured for some of them. Our numerical exploration is performed with a model which includes general relativity terms in the PN approximation, dynamical tides and handles potential collisions between pair of planets and with the star.
General relativity and dynamical tides were not included in the models of \cite{petrovich2014} and this may explain why eccentric and misaligned planets did not come out from their simulations.   We actually find that  
these two physical aspects significantly contribute to determine the final outcome of a P--P scattering event even if in different ways. General relativity, by changing the pericenter circulation frequency of the inner body, alters the secular evolution of a pair of planets close to the star. In particular, our modeling shows that  the secular free oscillations of the inner planet are damped while the forced eccentricity term is increased leading to an higher average eccentricity. This effect has implications  on the inward tidal migration which is determined by the orbital eccentricity modulating the pericenter distance. An important contribution of tides is that, after a period of chaotic evolution marked by mutual close encounters,  they may dynamically decouple the pair of planets surviving the period of instability. As a consequence, two planets can be found on stable  eccentric orbits even if one of them is the outcome of a merging event.  This decoupling is responsible for the formation of eccentric planets even when the planets begin their chaotic evolution very close to the star. 

By comparing the outcomes of our numerical modeling with the observed distribution of Jupiter--size planets we find a good match even if we do not have an explanation for the observed gap around 0.2 au unless some other dynamical mechanism prevents the planets to populate this region. However, this gap is suspected to be related to observational biases and for this reason we do not try to model it by introducing pre--determined and ad--hoc initial conditions. It is noteworthy that, according to our simulations, the majority of misaligned close--in giant planets should be the outcome of P--P scattering events where the inner planet is initially located beyond 0.2 au. If the inner planet is started on closer orbits, the surviving stable planets move near their star but on low inclination trajectories. Due to the scarcity of data on planet--star misalignment, we cannot predict the fraction of planets in close orbits that come from beyond 0.2 au. Additional observations are needed to derive a more robust estimate. 

Considering that the giant planets we deal with are located very close to the star and may be inflated either by stellar irradiation or because they are young  and in the early stages of their contraction, we have tested the relevance of the planet radius $r_p$  on the dynamical evolution of chaotic three--planet systems. The planetary radius is an important parameter as it determines the impact cross section and the strength of the tidal interaction. As an extreme case, we adopt a  radius of 1.5 $R_J$  for all the planets in the model and the same simulations performed for $r_p = R_J$ are run a second time with  1.5 $R_J$ for comparison. 
Only for system where the inner planet is very close to the star the chaotic evolution is significantly  influenced by the change in $r_p$. An increase in the number of circularized planets is observed while the overall eccentricity distribution shows slightly lower values. These effects, due to the increased strength of the tidal interaction, become progressively less relevant when the inner planet is moved outwards and are negligible when $a_1$ is located beyond 0.2 au. The increase in the cross section of the planets seems to be more relevant in terms of enhancing the tidal effects rather than reducing the number of scattering events which, within 1 au, is already small. 

A potential limitation of our modeling is the use of dynamical tides even when the eccentricity of the planet is small and static tides should come into play. However, this does not invalidate the results
since our model catches very well the dynamical evolution when the eccentricity of the planets is high and the most interesting dynamics occurs. The evolution predicted by dynamical tides is possibly too fast when the eccentricity becomes small but, at that stage of evolution, the orbital circularization would be simply due to tidal interaction and it would not be influenced by the presence of other planets or by dynamical scattering. This implies that planets which are circularized in our model on a short timescale would follow the same path under the slower static tides on longer time-spans.  This, for instance,  might be the case of HATS-52b \citep{henning2018}, a two Jupiter mass planet with an eccentricity of 0.24 and a semi--major axis of 0.025 au, or that of HAT-P-56b \citep{Huang2015}, a similar size planet with the same eccentricity but with a semi--major axis of 0.042 au.  

In most cases our modeling predicts the presence of an additional planet on an outer orbit. However, the second planet may be difficult to be detected by transits either because it is far from the star or because of the mutual inclination respect to the inner one due to the scattering process. However, in those cases where  also the second planet is left on a close orbit, it might be detected via TTVs because of its secular perturbations on the inner one. In this scenario, the models interpreting the TTV signal must take into account that very close to the star the secular theory is significantly affected by general relativity, as shown in  Figure~\ref{fig_sec1}  and Figure~\ref{fig_sec2}. 

When we compare our models with three initial planets to observations, we neglect those potential systems which were born with only two giant planets. They may  undergo P--P scattering as well and contribute to the final observed population. However, due to the high frequency with which collisions occur when the planets have gravitational encounters close to their star, we expect that these configurations preferentially lead to the formation of systems with a single giant planet moving on a close and low eccentricity orbit. By looking at the bottom left panel of Figure~\ref{fig_models}, we would guess that these single planets should contribute to the observed population of low eccentricity close--in Jupiters while those on eccentric/circularized and/or misaligned should originate from initially three--planets systems.

\begin{acknowledgements}
	We thank an anonymous referee and the Editor for useful comments and 
	suggestions. 
      ........
\end{acknowledgements}

\bibliographystyle{aa}
\bibliography{biblio}

\begin{thebibliography}{71}
\expandafter\ifx\csname natexlab\endcsname\relax\def\natexlab#1{#1}\fi

\bibitem[{{Alexander} \& {Armitage}(2009)}]{alexander2009}
{Alexander}, R.~D. \& {Armitage}, P.~J. 2009, \apj, 704, 989

\bibitem[{{Alexander} \& {Pascucci}(2012)}]{alexander2012}
{Alexander}, R.~D. \& {Pascucci}, I. 2012, \mnras, 422, L82

\bibitem[{{Alibert} {et~al.}(2005){Alibert}, {Mordasini}, {Benz}, \&
  {Winisdoerffer}}]{alibert2005}
{Alibert}, Y., {Mordasini}, C., {Benz}, W., \& {Winisdoerffer}, C. 2005, \aap,
  434, 343

\bibitem[{{Barker} \& {Ogilvie}(2009)}]{bar_ogil2009}
{Barker}, A.~J. \& {Ogilvie}, G.~I. 2009, \mnras, 395, 2268

\bibitem[{{Barnes}(2008)}]{barnes2008}
{Barnes}, R. 2008, {Dynamics of Multiple Planet Systems}, ed. J.~W. {Mason},
  177

\bibitem[{{Beaug{\'e}} \& {Nesvorn{\'y}}(2012)}]{beauge2012}
{Beaug{\'e}}, C. \& {Nesvorn{\'y}}, D. 2012, \apj, 751, 119

\bibitem[{{Chambers} {et~al.}(1996{\natexlab{a}}){Chambers}, {Wetherill}, \&
  {Boss}}]{chambers1996}
{Chambers}, J.~E., {Wetherill}, G.~W., \& {Boss}, A.~P. 1996{\natexlab{a}},
  \icarus, 119, 261

\bibitem[{{Chambers} {et~al.}(1996{\natexlab{b}}){Chambers}, {Wetherill}, \&
  {Boss}}]{chambers96}
{Chambers}, J.~E., {Wetherill}, G.~W., \& {Boss}, A.~P. 1996{\natexlab{b}},
  \icarus, 119, 261

\bibitem[{{Chatterjee} {et~al.}(2008){Chatterjee}, {Ford}, {Matsumura}, \&
  {Rasio}}]{CHATTERJEE08}
{Chatterjee}, S., {Ford}, E.~B., {Matsumura}, S., \& {Rasio}, F.~A. 2008, ApJ,
  686, 580

\bibitem[{{Damasso} {et~al.}(2015){Damasso}, {Esposito}, {Nascimbeni},
  {Desidera}, {Bonomo}, {Bieryla}, {Malavolta}, {Biazzo}, {Sozzetti}, {Covino},
  {Latham}, {Gandolfi}, {Rainer}, {Petrovich}, {Collins}, {Boccato}, {Claudi},
  {Cosentino}, {Gratton}, {Lanza}, {Maggio}, {Micela}, {Molinari}, {Pagano},
  {Piotto}, {Poretti}, {Smareglia}, {Di Fabrizio}, {Giacobbe}, {Gomez-Jimenez},
  {Murabito}, {Molinaro}, {Affer}, {Barbieri}, {Bedin}, {Benatti}, {Borsa},
  {Maldonado}, {Mancini}, {Scandariato}, {Southworth}, \& {Zanmar
  Sanchez}}]{damasso2015}
{Damasso}, M., {Esposito}, M., {Nascimbeni}, V., {et~al.} 2015, \aap, 581, L6

\bibitem[{{D'Angelo} {et~al.}(2006){D'Angelo}, {Lubow}, \& {Bate}}]{angelo2006}
{D'Angelo}, G., {Lubow}, S.~H., \& {Bate}, M.~R. 2006, \apj, 652, 1698

\bibitem[{{Donnison}(2006)}]{donnison2006}
{Donnison}, J.~R. 2006, \mnras, 369, 1267

\bibitem[{{Duffell} \& {Chiang}(2015)}]{duffell2015}
{Duffell}, P.~C. \& {Chiang}, E. 2015, \apj, 812, 94

\bibitem[{{Emsenhuber} \& {Mordasini}(2019)}]{emme19}
{Emsenhuber}, A. \& {Mordasini}, C. 2019, in press.

\bibitem[{{Ercolano} \& {Rosotti}(2015)}]{ercolano2015}
{Ercolano}, B. \& {Rosotti}, G. 2015, \mnras, 450, 3008

\bibitem[{{Fabrycky} \& {Tremaine}(2007)}]{fabry2007}
{Fabrycky}, D. \& {Tremaine}, S. 2007, \apj, 669, 1298

\bibitem[{{Fortney} {et~al.}(2007){Fortney}, {Marley}, \&
  {Barnes}}]{fortney2007}
{Fortney}, J.~J., {Marley}, M.~S., \& {Barnes}, J.~W. 2007, \apj, 659, 1661

\bibitem[{{Gladman}(1993)}]{gladman1993}
{Gladman}, B. 1993, \icarus, 106, 247

\bibitem[{{Goldreich} \& {Sari}(2003)}]{sari2003}
{Goldreich}, P. \& {Sari}, R. 2003, \apj, 585, 1024

\bibitem[{{Goldreich} \& {Tremaine}(1980)}]{gold1980}
{Goldreich}, P. \& {Tremaine}, S. 1980, \apj, 241, 425

\bibitem[{{Hasegawa} \& {Pudritz}(2012)}]{hasegawa2012}
{Hasegawa}, Y. \& {Pudritz}, R.~E. 2012, \apj, 760, 117

\bibitem[{Henning {et~al.}(2018)Henning, Mancini, Sarkis, Bakos, Hartman,
  Bayliss, Bento, Bhatti, Brahm, Ciceri, Csubry, de~Val-Borro, Espinoza,
  Fulton, Howard, Isaacson, Jordán, Marcy, Penev, Rabus, Suc, Tan, Tinney,
  Wright, Zhou, Durkan, Lazar, Papp, \& Sari}]{henning2018}
Henning, T., Mancini, L., Sarkis, P., {et~al.} 2018, The Astronomical Journal,
  155, 79

\bibitem[{{Huang} {et~al.}(2015){Huang}, {Hartman}, {Bakos}, {Penev}, {Bhatti},
  {Bieryla}, {de Val-Borro}, {Latham}, {Buchhave}, {Csubry}, {Kov{\'a}cs},
  {B{\'e}ky}, {Falco}, {Berlind}, {Calkins}, {Esquerdo}, {L{\'a}z{\'a}r},
  {Papp}, \& {S{\'a}ri}}]{Huang2015}
{Huang}, C.~X., {Hartman}, J.~D., {Bakos}, G.~{\'A}., {et~al.} 2015, \aj, 150,
  85

\bibitem[{{Ivanov} \& {Papaloizou}(2004)}]{ivanov2004}
{Ivanov}, P.~B. \& {Papaloizou}, J.~C.~B. 2004, \mnras, 347, 437

\bibitem[{{Ivanov} \& {Papaloizou}(2007)}]{ivanov2007}
{Ivanov}, P.~B. \& {Papaloizou}, J.~C.~B. 2007, \mnras, 376, 682

\bibitem[{{Ivanov} \& {Papaloizou}(2011)}]{ivanov2011}
{Ivanov}, P.~B. \& {Papaloizou}, J.~C.~B. 2011, Celestial Mechanics and
  Dynamical Astronomy, 111, 51

\bibitem[{{Juri{\'c}} \& {Tremaine}(2008)}]{juritre08}
{Juri{\'c}}, M. \& {Tremaine}, S. 2008, ApJ, 686, 603

\bibitem[{{Kidder}(1995)}]{kidder1995}
{Kidder}, L.~E. 1995, \prd, 52, 821

\bibitem[{{Kley} \& {Nelson}(2012)}]{kleynelson2012}
{Kley}, W. \& {Nelson}, R.~P. 2012, \araa, 50, 211

\bibitem[{{Kokubo} {et~al.}(1998){Kokubo}, {Yoshinaga}, \& {Makino}}]{kokubo98}
{Kokubo}, E., {Yoshinaga}, K., \& {Makino}, J. 1998, \mnras, 297, 1067

\bibitem[{{Lai}(1997)}]{lai97}
{Lai}, D. 1997, \apj, 490, 847

\bibitem[{{Lee} \& {Ostriker}(1986)}]{Lee1986}
{Lee}, H.~M. \& {Ostriker}, J.~P. 1986, \apj, 310, 176

\bibitem[{{Lee} \& {Peale}(2002)}]{Leepeale2002}
{Lee}, M.~H. \& {Peale}, S.~J. 2002, \apj, 567, 596

\bibitem[{{Lega} {et~al.}(2013{\natexlab{a}}){Lega}, {Morbidelli}, \&
  {Nesvorn{\'y}}}]{lega2013}
{Lega}, E., {Morbidelli}, A., \& {Nesvorn{\'y}}, D. 2013{\natexlab{a}}, \mnras,
  431, 3494

\bibitem[{{Lega} {et~al.}(2013{\natexlab{b}}){Lega}, {Morbidelli}, \&
  {Nesvorn{\'y}}}]{legamorb2013}
{Lega}, E., {Morbidelli}, A., \& {Nesvorn{\'y}}, D. 2013{\natexlab{b}}, \mnras,
  431, 3494

\bibitem[{{Libert} \& {Henrard}(2005)}]{henrard2005}
{Libert}, A.-S. \& {Henrard}, J. 2005, Celestial Mechanics and Dynamical
  Astronomy, 93, 187

\bibitem[{{Mardling}(1995{\natexlab{a}})}]{mardling1995b}
{Mardling}, R.~A. 1995{\natexlab{a}}, \apj, 450, 722

\bibitem[{{Mardling}(1995{\natexlab{b}})}]{mardling1995a}
{Mardling}, R.~A. 1995{\natexlab{b}}, \apj, 450, 732

\bibitem[{{Marzari}(2010)}]{marzari10}
{Marzari}, F. 2010, Formation and Evolution of Exoplanets, R. Barnes Editor,
  WILEY-VCH, 223

\bibitem[{{Marzari}(2014)}]{marzari2014}
{Marzari}, F. 2014, \mnras, 442, 1110

\bibitem[{{Marzari} {et~al.}(2010{\natexlab{a}}){Marzari}, {Baruteau}, \&
  {Scholl}}]{marza-barute2010}
{Marzari}, F., {Baruteau}, C., \& {Scholl}, H. 2010{\natexlab{a}}, \aap, 514,
  L4

\bibitem[{{Marzari} {et~al.}(2010{\natexlab{b}}){Marzari}, {Baruteau}, \&
  {Scholl}}]{marbaru2010}
{Marzari}, F., {Baruteau}, C., \& {Scholl}, H. 2010{\natexlab{b}}, \aap, 514,
  L4

\bibitem[{{Marzari} \& {Nelson}(2009)}]{marnel2009}
{Marzari}, F. \& {Nelson}, A.~F. 2009, \apj, 705, 1575

\bibitem[{{Marzari} \& {Picogna}(2013)}]{marpi2013}
{Marzari}, F. \& {Picogna}, G. 2013, \aap, 550, A64

\bibitem[{{Marzari} \& {Weidenschilling}(2002{\natexlab{a}})}]{MW02}
{Marzari}, F. \& {Weidenschilling}, S.~J. 2002{\natexlab{a}}, Icarus, 156, 570

\bibitem[{{Marzari} \& {Weidenschilling}(2002{\natexlab{b}})}]{marweiden2002}
{Marzari}, F. \& {Weidenschilling}, S.~J. 2002{\natexlab{b}}, \icarus, 156, 570

\bibitem[{{Masset} \& {Snellgrove}(2001)}]{masset2001}
{Masset}, F. \& {Snellgrove}, M. 2001, \mnras, 320, L55

\bibitem[{{Mathis}(2015)}]{mathis2015}
{Mathis}, S. 2015, \aap, 580, L3

\bibitem[{{Matsuyama} {et~al.}(2003){Matsuyama}, {Johnstone}, \&
  {Murray}}]{matsuyama2003}
{Matsuyama}, I., {Johnstone}, D., \& {Murray}, N. 2003, \apjl, 585, L143

\bibitem[{{Misner} {et~al.}(1973){Misner}, {Thorne}, \& {Wheeler}}]{misner73}
{Misner}, C.~W., {Thorne}, K.~S., \& {Wheeler}, J.~A. 1973, {Gravitation}

\bibitem[{{Mordasini} {et~al.}(2012){Mordasini}, {Alibert}, {Georgy},
  {Dittkrist}, {Klahr}, \& {Henning}}]{mordasini2012}
{Mordasini}, C., {Alibert}, Y., {Georgy}, C., {et~al.} 2012, \aap, 547, A112

\bibitem[{{Murray} \& {Dermott}(1999)}]{murray-dermottSS}
{Murray}, C.~D. \& {Dermott}, S.~F. 1999, {Solar system dynamics}

\bibitem[{{Mustill} {et~al.}(2014){Mustill}, {Veras}, \& {Villaver}}]{must2014}
{Mustill}, A.~J., {Veras}, D., \& {Villaver}, E. 2014, \mnras, 437, 1404

\bibitem[{{Nagasawa} \& {Ida}(2011)}]{naga11}
{Nagasawa}, M. \& {Ida}, S. 2011, ApJ, 742, 72

\bibitem[{{Nagasawa} {et~al.}(2008){Nagasawa}, {Ida}, \& {Bessho}}]{naga08}
{Nagasawa}, M., {Ida}, S., \& {Bessho}, T. 2008, ApJ, 678, 498

\bibitem[{{Namouni}(2007)}]{namouni07}
{Namouni}, F. 2007, in Lecture Notes in Physics, Berlin Springer Verlag, Vol.
  729, Lecture Notes in Physics, Berlin Springer Verlag, ed. D.~{Benest},
  C.~{Froeschle}, \& E.~{Lega}, 233

\bibitem[{{Naoz}(2016)}]{naoz2016}
{Naoz}, S. 2016, \araa, 54, 441

\bibitem[{{Naoz} {et~al.}(2011){Naoz}, {Farr}, {Lithwick}, {Rasio}, \&
  {Teyssandier}}]{naoz2011}
{Naoz}, S., {Farr}, W.~M., {Lithwick}, Y., {Rasio}, F.~A., \& {Teyssandier}, J.
  2011, \nat, 473, 187

\bibitem[{{Nelson} {et~al.}(2017){Nelson}, {Ford}, \& {Rasio}}]{nelson2017}
{Nelson}, B.~E., {Ford}, E.~B., \& {Rasio}, F.~A. 2017, \aj, 154, 106

\bibitem[{{Ogilvie}(2013)}]{ogilvie2013}
{Ogilvie}, G.~I. 2013, \mnras, 429, 613

\bibitem[{{Petrovich} \& {Tremaine}(2016)}]{petrovich_secular_2016}
{Petrovich}, C. \& {Tremaine}, S. 2016, \apj, 829, 132

\bibitem[{{Petrovich} {et~al.}(2014){Petrovich}, {Tremaine}, \&
  {Rafikov}}]{petrovich2014}
{Petrovich}, C., {Tremaine}, S., \& {Rafikov}, R. 2014, \apj, 786, 101

\bibitem[{{Rasio} \& {Ford}(1996)}]{rasio-ford96}
{Rasio}, F.~A. \& {Ford}, E.~B. 1996, Science, 274, 954

\bibitem[{{Raymond} {et~al.}(2009){Raymond}, {Armitage}, \& {Gorelick}}]{ray09}
{Raymond}, S.~N., {Armitage}, P.~J., \& {Gorelick}, N. 2009, ApJL, 699, L88

\bibitem[{{Raymond} {et~al.}(2008){Raymond}, {Barnes}, {Armitage}, \&
  {Gorelick}}]{ray08}
{Raymond}, S.~N., {Barnes}, R., {Armitage}, P.~J., \& {Gorelick}, N. 2008,
  ApJL, 687, L107

\bibitem[{{Schneider} {et~al.}(2011){Schneider}, {Dedieu}, {Le Sidaner},
  {Savalle}, \& {Zolotukhin}}]{enciclo}
{Schneider}, J., {Dedieu}, C., {Le Sidaner}, P., {Savalle}, R., \&
  {Zolotukhin}, I. 2011, \aap, 532, A79

\bibitem[{{Shara} {et~al.}(2016){Shara}, {Hurley}, \& {Mardling}}]{shara2016}
{Shara}, M.~M., {Hurley}, J.~R., \& {Mardling}, R.~A. 2016, \apj, 816, 59

\bibitem[{{Veras} \& {Mustill}(2013)}]{veras2013}
{Veras}, D. \& {Mustill}, A.~J. 2013, \mnras, 434, L11

\bibitem[{{Weidenschilling} \& {Marzari}(1996)}]{weiden-marza96}
{Weidenschilling}, S.~J. \& {Marzari}, F. 1996, Nature, 384, 619

\bibitem[{Wise \& Dodson-Robinson(2018)}]{wise2018}
Wise, A. \& Dodson-Robinson, S. 2018, Research Notes of the AAS, 2, 29

\bibitem[{{Wise} \& {Dodson-Robinson}(2018)}]{wise2018-2}
{Wise}, A.~W. \& {Dodson-Robinson}, S.~E. 2018, \apj, 855, 145

\end{thebibliography}


\end{document}